\documentstyle[twocolumn,aps,floats,amssymb,epsfig]{revtex}

\begin{document}

\def\d{{\rm d}}
\def\e{{\rm e}}
\def\i{{\rm i}}
\def\O{{\rm O}}
\def\half{\mbox{$\frac12$}}
\def\eref#1{(\protect\ref{#1})}
\def\etal{{\it{}~et~al.}}
\def\Li{\mathop{\rm Li}}
\def\av#1{\left\langle#1\right\rangle}
\def\set#1{\left\lbrace#1\right\rbrace}
\def\stirling#1#2{\Bigl\lbrace{#1\atop#2}\Bigr\rbrace}

\draft
\tolerance = 10000

\renewcommand{\topfraction}{0.9}
\renewcommand{\textfraction}{0.1}
\renewcommand{\floatpagefraction}{0.9}
\setlength{\tabcolsep}{4pt}

%Fixing abstract in twocolumn mode
\twocolumn[\hsize\textwidth\columnwidth\hsize\csname @twocolumnfalse\endcsname

\title{Are randomly grown graphs really random?}
\author{Duncan S. Callaway$^{1}$, John E. Hopcroft$^{2}$,
Jon M. Kleinberg$^{2}$, M. E. J. Newman$^{3,4}$,
and Steven H. Strogatz$^{1,4}$}
\address{$^1$Department of Theoretical and Applied Mechanics,
Cornell University, Ithaca NY 14853--1503}
\address{$^2$Department of Computer Science, Cornell University,
Ithaca NY 14853}
\address{$^3$Santa Fe Institute, 1399 Hyde Park Road, Santa Fe NM 87501}
\address{$^4$Center for Applied Mathematics, Cornell University,
Ithaca NY 14853--3801}
\maketitle

\begin{abstract}
  We analyze a minimal model of a growing network. At each time step, a new
  vertex is added; then, with probability $\delta$, two vertices are chosen
  uniformly at random and joined by an undirected edge.  This process is
  repeated for $t$ time steps. In the limit of large $t$, the resulting
  graph displays surprisingly rich characteristics.  In particular, a giant
  component emerges in an infinite--order phase transition at $\delta =
  1/8.$ At the transition, the average component size jumps discontinuously
  but remains finite. In contrast, a static random graph with the same
  degree distribution exhibits a second--order phase transition at $\delta
  = 1/4$, and the average component size diverges there. These dramatic
  differences between grown and static random graphs stem from a positive
  correlation between the degrees of connected vertices in the grown
  graph---older vertices tend to have higher degree, and to link with other
  high--degree vertices, merely by virtue of their age. We conclude that
  grown graphs, however randomly they are constructed, are fundamentally
  different from their static random graph counterparts.
\end{abstract}

%\pacs{}
\vspace{1cm}

%Fixing abstract in twocolumn mode
]

\section{Introduction}
Many networks grow over time.  New pages and links are added to the World
Wide Web every day, while networks like the power grid, the Internet
backbone, and social networks change on slower time-scales.  Even naturally
occurring networks such as food webs and biochemical networks evolve.

In the last few years, physicists, mathematicians, and computer scientists
have begun to explore the structural implications of network growth, using
techniques from statistical mechanics, graph theory, and computer
simulation~\cite{bara99a,bara99b,albe00a,brod00,doro00a,doro00b,doro00c,krap00a,krap00b,krap01,kuma00,tadi01}.
Much of this research has been stimulated by recent discoveries about the
structure of the World Wide Web, metabolic networks, collaboration
networks, the Internet, food webs, and other complex
networks~\cite{brod00,redn98,falo99,albe99,hube99,jeon00,stro01}.  

Among the many properties of these networks that have been studied, one
that has assumed particular importance is the degree distribution.  The
degree of a vertex in a network is the number of other vertices to which
it is connected.  Many real-world networks are found to have highly
skewed degree distributions, such that most vertices have only a small
number of connections to others, but there are a few, like Yahoo and CNN
in the Web, or ATP and carbon dioxide in biochemical reaction networks,
which are very highly connected.  If we define $p_k$ to be the probability
that a randomly chosen vertex has $k$ neighbors, it turns out that $p_k$
often has either a power-law tail as a function of $k$ (indicating that
there is no characteristic scale for the degree), or a power-law tail
truncated by an exponential
cutoff~\cite{brod00,redn98,falo99,albe99,hube99,jeon00,amar00}.  These
distributions are quite different from the single-scale Poisson
distribution seen in traditional random graph models of
networks~\cite{erdo59_61,boll85}.

One theoretical challenge has been to explain the origin of these observed
degree distributions.  Barab\'asi and co-workers~\cite{bara99a,bara99b}
have emphasized the key role played by network growth.  They showed that a
power-law degree distribution emerges naturally from a stochastic growth
process in which new vertices link to existing ones with a probability
proportional to the degree of the target vertex.  More refined variants of
this preferential attachment process allow for aging of vertices, rewiring
of edges, and nonlinear attachment probability, with power laws or
truncated power laws emerging for a wide range of
assumptions~\cite{doro00a,doro00b,doro00c,krap00a,krap00b,krap01}.
Kumar\etal~\cite{kuma00} have concurrently proposed a model in which a
local copying process for edges leads to a type of preferential attachment
phenomenon as well.

As in these studies, we consider the role of system growth on network
structure.  However, our purpose is somewhat different.  Rather than
seeking to explain an observed feature of real-world networks, such as the
degree distribution, we focus on a minimal model of network growth and
compare its properties to those of the familiar random graph.  We do not
claim that our model is an accurate reflection of any particular real-world
system, but we find that studying a model that exhibits network growth in
the absence of other complicating features leads to several useful
insights.  In addition, the model turns out to have some interesting
mathematical properties, as we will show.

Among other things, we solve for the distribution of the sizes of
components (connected sets of vertices), a distribution that
has not been studied in previous growth models, largely because most of
them produce only one huge, connected component.  We find that the model
exhibits a phase transition at which a giant component forms---a component
whose size scales linearly with system size.  In this respect our networks
resemble traditional random graphs~\cite{erdo59_61,boll85}, but they differ
from random graphs in many other ways.  For example, the mean component
size is different both quantitatively and also qualitatively, having no
divergence at the phase transition.  The position of the phase transition
is different as well, and the transition itself appears to be infinite
order rather than second order.  There are thus a number of features, both
local and global, by which the grown graph can be distinguished from a
static one.

In a certain sense, therefore, it seems that a randomly grown network is
not really random.

\section{The model}
\label{model}
Our model is very simple.  At each time step, a new vertex is added.  Then,
with probability~$\delta$, two vertices are chosen uniformly at random and
joined by an undirected edge.  Our goal is to understand the statistical
properties of the network in the limit of large time~$t$.

This model differs from preferential attachment models in two
important ways.  First, new edges are introduced between randomly
chosen pairs of vertices, with no preference given to high degree
vertices.  Second, new vertices do not necessarily attach to a
pre-existing vertex when they enter the network.  In other words,
there is no guarantee that a new vertex will have an edge emanating
from it.  As a result the graphs generated by our model usually
contain isolated vertices, along with components of various sizes,
whereas the preferential attachment models typically generate graphs in
which all vertices are connected in a single component.

\section{Degree distribution}
\label{degree}
We begin by calculating the distribution of vertex degrees in our model.
For concreteness, we choose an initial condition for the graph in which
there is a single isolated vertex and no edges, although the asymptotic
behavior at long times does not depend on this initial condition.

At time $t$ there will be $t$ vertices and on average $\delta t$ edges.
Let $d_k(t)$ be the expected number of vertices with degree~$k$ at
time~$t$.  The number of isolated vertices, $d_0(t)$, will increase by one
at each time step, but decrease on average by $2\delta d_0(t)/t$, the
probability that a degree zero vertex is randomly chosen as one of the ends
of a new edge.  Thus
\begin{equation}
d_0(t+1) = d_0(t) + 1 - 2\delta\frac{d_0(t)}{t}.
\label{degree1}
\end{equation}
Similarly, the expected number of degree~$k$ vertices ($k>0$) will increase
on average by an amount proportional to the probability that a degree~$k-1$
vertex is chosen for attachment by a new edge, and decrease by an amount
proportional to the probability that a degree~$k$ vertex is chosen.  This
gives
\begin{equation}
d_k(t+1) = d_k(t) + 2\delta\frac{d_{k-1}(t)}{t} - 2\delta\frac{d_k(t)}{t}.
\label{degree2}
\end{equation}
Note that these equations neglect the possibility that an edge links a
vertex to itself.  This means the equations are only approximate at short
times, but they become exact in the limit $t\to\infty$ because the
probability that any vertex is chosen twice decreases like $t^{-2}$.

For large $t$,  numerical simulations show that solutions of these
equations grow linearly in time: $d_k(t)\sim p_kt$.  Seeking solutions of
this form, we find that $p_0 = 1/(1+2\delta)$, and $p_k =
(2\delta/(1+2\delta))^kp_0$ for $k>0$.  Thus, in general, the probability
of a randomly chosen vertex having degree $k$ is

\begin{equation}
p_k = \frac{(2\delta)^k}{(1+2\delta)^{k+1}}.
\label{p_k_soln}
\end{equation}

In other words, the randomly grown network has an exponential degree
distribution.  This result will become important shortly.

\section{Critical behavior}
\label{crit_sect}
In this section we establish that the grown graph displays a phase transition
for finite $\delta$ at which a giant component forms, and study the
critical behavior of the system in the vicinity of this transition.

\begin{figure}
\begin{center}
\psfig{figure=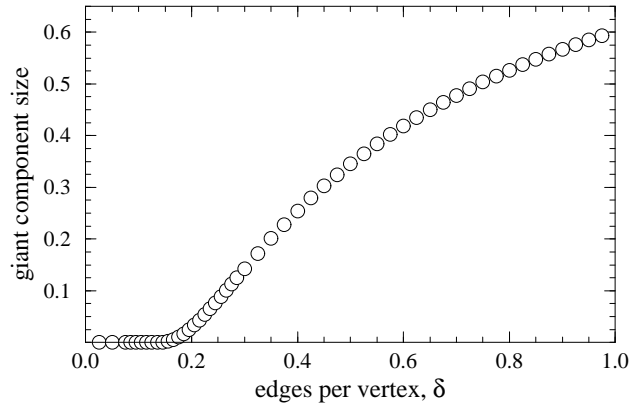,width=3.3in}
\caption{Giant component size $S$ in the randomly grown graph, as a
  function of $\delta$.  Here $S$ is defined as the number of vertices in
  the largest component, divided by the system size~$t$.  Results are
  obtained by simulating the growing graph for $1.6\times 10^7$ time steps,
  with the number of edges assigned by a Bernoulli distribution of
  mean~$\delta$, i.e.,~one edge is introduced per time step with
  probability $\delta$; otherwise no edges are introduced.  Component sizes
  were calculated by depth-first search.  The results shown are an average
  over 25 repetitions of the calculation.}
\label{gc_notheory}
\end{center}
\end{figure}

\subsection{Size of the giant component}
Fig.~\ref{gc_notheory} shows the average size $S$ of the largest component
in simulations of our model for a range of values of the
parameter~$\delta$, as a fraction of the total system size.  From the
figure, it appears that a giant component forms somewhere between
$\delta=0.1$ and $\delta=0.2$, although no discontinuity is apparent in
$S(\delta)$ or in its derivative.  The smoothness of this transition sets
the growing graph apart from random graphs, for which there is known to be
a discontinuity in the first derivative of the giant component size at the
transition.

To address the difference between static and growing graphs analytically,
let $N_k(t)$ be the expected number of components of size~$k$.  At each
time step, one isolated vertex (i.e.,~a component of size one) is added to
the graph.  At the same time, $2\delta N_1(t)/t$ vertices will on average
be chosen for attachment and thereby leave the set of isolated vertices.
Thus $N_1(t)$ obeys
\begin{equation}
N_1(t+1)=N_1(t) + 1 - 2\delta \frac{N_1(t)}{t}.
\label{N1_eqn}
\end{equation}

Components of size $k>1$ are gained when vertices belonging to separate
components whose size sums to $k$ are connected by a new edge.  Components
of size $k>1$ are lost when a vertex within a $k$-sized component is chosen
for attachment.  Thus the number of components of size $k>1$ satisfies
\begin{eqnarray}
N_k(t+1) &=& N_k(t) +
\delta \sum_{j=1}^{k-1}\frac{jN_j(t)}{t}\,\frac{(k-j)N_{k-j}(t)}{t}\nonumber\\
& & \null - 2\delta \frac{kN_k(t)}{t}.
\label{Nk_eqn}
\end{eqnarray}

As with Eqs.~\eref{degree1} and~\eref{degree2} for the degree distribution,
these equations are approximate for small system sizes because we have
neglected the possibility that both ends of an edge fall within the same
component.  This probability tends to zero as system size becomes large,
and hence the equations become exact in the limit $t\to\infty$.
Equivalently, there is a negligible probability of closed loops within any
component of fixed size~$k$, as $t\to\infty$.  Of course, there can be
closed loops in the giant component, if one exists.  Thus,
Eqs.~\eref{N1_eqn} and~\eref{Nk_eqn} hold only for the finite components in
the graph, a fact which we exploit below.

Seeking solutions to Eqs.~\eref{N1_eqn} and~\eref{Nk_eqn} of the form
$N_k(t)=a_kt$, where $a_k$ is the steady-state solution of the component
size distribution, we find that
\begin{eqnarray}
\label{a1_eqn}
a_1 &=& \frac{1}{1+2\delta}\\
\label{ak_eqn}
a_k &=& \frac{\delta}{1+2k\delta} \sum_{j=1}^{k-1} j(k-j)a_ja_{k-j}.
\end{eqnarray}
For any given $k$, the coefficient $a_k$ can be calculated from these
equations by explicit iteration.  We however will take a different approach
here and derive closed-form results by defining a generating function $g(x)$
for the distribution of component sizes:
\begin{equation}
g(x)=\sum_{k=1}^{\infty} b_k x^k,
\label{gf_def_eqn}
\end{equation}
where
\begin{equation}
b_k=ka_k.
\label{bk_def_eqn}
\end{equation}
The coefficient $b_k$ has a simple interpretation: it is the probability that
a randomly chosen vertex belongs to a finite component containing exactly $k$
vertices.

If we multiply both sides of Eqs.~\eref{a1_eqn} and~\eref{ak_eqn} by $kx^k$
and sum over~$k$, we find that $g(x)$ satisfies the differential equation
\begin{equation}
\label{gf_ode_eqn}
g = -2 \delta x g' + 2 \delta x g g' + x,
\end{equation}
where $g' = \d g/\d x$.  Rearranging for $g'$ then yields
\begin{equation}
g' = {1\over2\delta}\,\left[\frac{1-g/x}{1-g}\right].
\label{gprime_eqn}
\end{equation}

The generating function $g(x)$ provides a convenient way to determine the
size $S$ of the giant component.  We observe that $g(1)=\sum_{k=1}^{\infty}
b_k$, and hence that $g(1)$ is the probability that a randomly chosen
vertex will belong to some component of finite size (since, as we mentioned
above, the quantities $N_k$ and hence also $b_k$ represent only the
finite-sized components).  When no giant component exists, this probability is
exactly~1, but if there is a giant component, then $g(1)<1$ and the size of
the giant component is

\begin{eqnarray}
S=1-g(1).
\label{gceqn}
\end{eqnarray}

In the absence of an analytic solution for Eq.~\eref{gprime_eqn} we
evaluate $S$ numerically by integrating~\eref{gprime_eqn} using the initial
condition $(x,g(x))=(x_0,x_0/(1+2\delta))$ for small $x_0$.  (We find that
$x_0=10^{-6}$ gives sufficient accuracy.)  The resulting value of $S$ is
shown as a solid line in Fig.~\ref{gc_fig}, and is in good agreement with
the data from the direct simulations of the model (circles), suggesting,
among other things, that it was a reasonable approximation to neglect
closed loops in finite-sized components, as we claimed above.

\subsection{Comparison with a static random graph}
We now compare our results for the grown network with the properties of an
ordinary static random graph, in which edges are added to a pre-existing
complete set of vertices and no new vertices are ever added.  The standard
example of such a static graph is the so-called $G_{n,p}$ model of
Erd\H{o}s and R\'enyi~\cite{erdo59_61}.  This model however does not
provide an ideal benchmark, since the degree distribution for $G_{n,p}$ is
Poisson whereas the distribution for our networks is exponential, as we
showed in Section~\ref{degree}.  Fortunately, it is possible to construct a
random graph that has an exponential degree distribution (or any other
distribution we desire) using the construction given by Molloy and
Reed~\cite{moll95,moll98}, which works as follows.
\begin{enumerate} 
\item Create a set of $N$ vertices, indexed by $i = 1,2,\ldots N$, whose
  degree $k_i$ is assigned from the distribution of interest.
\item Form a list $L$ that contains $k_i$ distinct copies of each
  vertex~$i$.
\item Choose a random matching of pairs of elements of $L$ to create an
  edge list.
\end{enumerate} 

\begin{figure}
\begin{center}
\psfig{figure=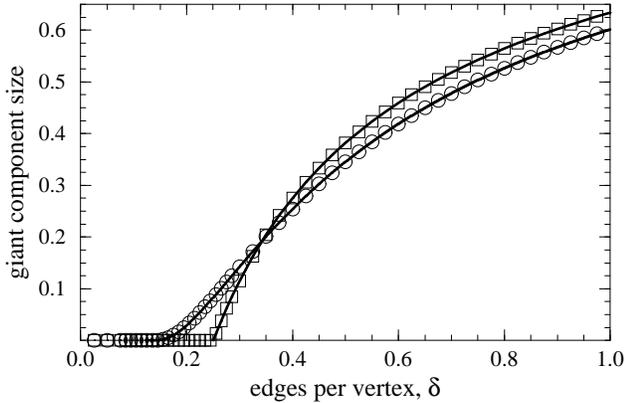,width=3.3in}
\caption{Size $S$ of the largest component for the randomly grown network
  (circles), and for a static random graph with same degree distribution
  (squares).  Points are results from numerical simulations and the solid
  lines are theoretical results from Eq.~\eref{gceqn} and
  Ref.~\protect\onlinecite{newm00}.  The grown graph was simulated for
  1.6$\times 10^7$ time steps, starting from a single site.}
\label{gc_fig}
\end{center}
\end{figure}

As with the model of Erd\H{o}s and R\'enyi, this model exhibits a
distribution of component sizes and a phase transition at which a giant
component of size $O(N)$ appears~\cite{moll95,newm00}.  In
Fig.~\ref{gc_fig} (squares) we show numerical results for the size of this
giant component for a static random graph with degree distribution
identical to that of our grown graph, i.e.,~conforming to
Eq.~\eref{p_k_soln}.  The size of the giant component can also be
calculated exactly in the limit of large graph size using results due to
Molloy and Reed~\cite{moll98}, or equivalently using the generating
function formalism of Newman\etal~\cite{newm00}.  The result is
\begin{equation}
S_{\text{static}} = \biggl\lbrace \begin{array}{ll}
  0,\qquad                                    & \delta\le\frac14 \\
  1-1/(\delta+\sqrt{\delta^2+2\delta}),\qquad & \delta>\frac14,
\end{array}
\label{s_random}
\end{equation}
which is shown as a solid line in Fig.~\ref{gc_fig}.

Figure~\ref{gc_fig} shows that there is a marked discrepancy between the
size of the giant component in the static and grown cases.  In the
following sections we show analytically that this is indeed the case
by locating critical values of $\delta$ at which the giant components
form.

\subsection{Average component size and position of the phase transition}

For the static graph with the same exponential degree distribution as our
grown graph, Eq.~\eref{s_random} shows that the size $S_{\text{static}}$ of
the giant component tends to zero continuously and vanishes at $\delta_c =
\frac14$.  For the grown model, we do not have an analogous closed-form
result for $S(\delta)$.  However, we can still find the value of $\delta_c$
by considering the average size $\av{s}$ of the finite components, which is
given in terms of the generating function $g(x)$ by
\begin{equation}
\av{s}=\frac{g'(1)}{g(1)}.
\label{av_s_eqn}
\end{equation}
To locate the transition, we examine the behavior of $g'(1)$, using
Eq.~\eref{gprime_eqn}.

For values of $\delta$ where the giant component exists, we have $g(1)\ne
1$ and, setting $x=1$ in Eq.~\eref{gprime_eqn}, we find that
\begin{equation}
g'(1)=\frac{1}{2\delta}, \qquad\mbox{when $g(1)\ne1$}.
\label{hi_soln}
\end{equation}
This equation holds for all $\delta>\delta_c$, where $\delta_c$ still
remains to be determined.  Conversely, if $\delta<\delta_c$, the giant
component does not exist and $g(1)=1$, in which case both the numerator and
denominator of Eq.~\eref{gprime_eqn} approach zero as $x\to1$.  Applying
L'Hopital's rule we then derive a quadratic equation for $g'(1)$, whose
solution is
\begin{equation}
g'(1) = \frac{1\pm\sqrt{1-8\delta}}{4\delta}, \qquad\mbox{when $g(1)=1$}.
\label{low_soln}
\end{equation}
This solution is only valid for $0\le\delta\le\frac18$.

Thus for all $\delta>\frac18$ we have only a single solution~\eref{hi_soln}
for $g'(1)$, which necessarily means that a giant component exists.  For
$\delta\le\frac18$, we have three solutions, one of which~\eref{hi_soln}
implies the existence of a giant component while the other
two~\eref{low_soln} do not.  Thus the phase transition, if there is one,
must occur at $\delta_c\le\frac18$.

If we make the further observation that in the limit $\delta\to0$ all
components have size~1, it is clear that the correct solution for $g'(1)$
in this limit is Eq.~\eref{low_soln} with the negative sign.  In the
absence of any non-analytic behavior in the solution of
Eq.~\eref{gprime_eqn} other than at $\delta=\frac18$, we then conclude that
in fact this branch is the correct solution for all $0\le\delta\le\frac18$,
and hence that
\begin{equation}
\delta_c = \mbox{$\frac18$}.
\end{equation}
This is clearly different from the $\delta_c=\frac14$ of the static model,
and agrees qualitatively with what we observe in Fig.~\ref{gc_fig}.

In summary,
\begin{equation}
g'(1) = \biggl\lbrace \begin{array}{ll}
(1-\sqrt{1-8\delta})/4\delta,\qquad & \delta\le\frac18\\
1/2\delta,\qquad & \delta >\frac18,
\end{array}
\label{gprime_soln}
\end{equation}
which implies that $g'(1)$ jumps discontinuously from 2 to 4 as $\delta$
passes through $\delta=\frac18$ and hence that the average component size
$\av{s}$ also jumps from 2 to 4 at the transition.

\begin{figure}
\begin{center}
\psfig{figure=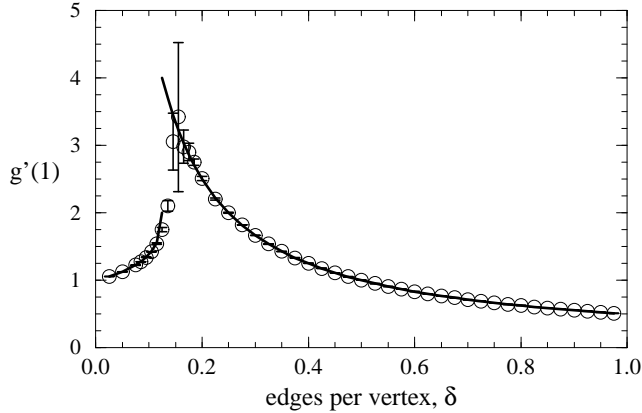,width=3.3in}
\end{center}
\caption{Discontinuous behavior of $g'(1)$ for the growing graph.  The solid
  line is the theoretical prediction from Eq.~\eref{gprime_soln} and the
  open circles are data from simulations of the growing network for
  1.6$\times 10^7$ time steps (averaged over many runs).}
\label{gprime_fig}
\end{figure}

In Fig.~\ref{gprime_fig} we compare our analytic results for $g'(1)$
with direct simulations of the model.  The predicted discontinuity is
clearly visible, although, as is typical with simulations near critical
points, the numerical errors are large in the region of the transition.

\subsection{Infinite-order transition}
The phase transitions of grown and static random networks differ in more
than just their location.  The random graph undergoes a second-order phase
transition ($S_{\text{static}}$ is continuous but its first derivative
with respect to $\delta$ is discontinuous at $\delta=\frac14$), whereas
the transition for the growing graph is of at least third order ($S$ and
its first derivative appear continuous at $\delta=\frac18$ from inspection
of Fig.~\ref{gc_notheory}).

\begin{figure}
\begin{center}
\psfig{figure=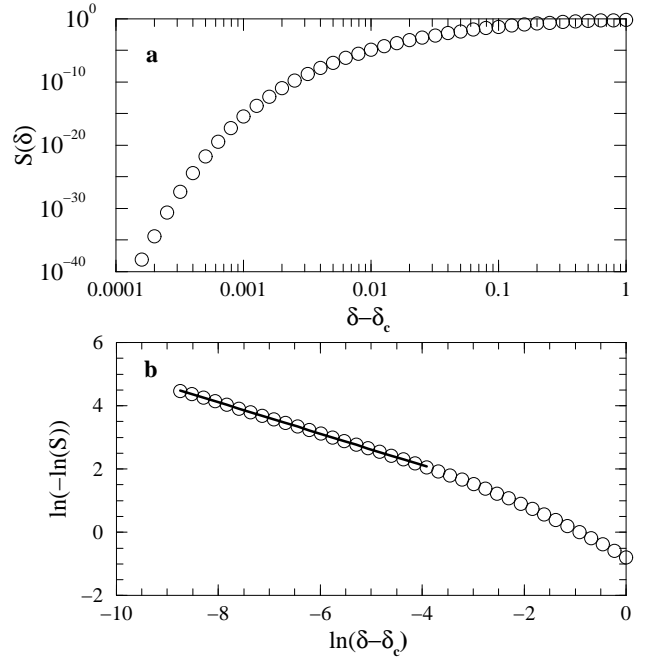,width=3.3in}
\end{center}
\caption{Giant component size $S(\delta)$ near the phase transition, from
  numerical integration of Eq.~\eref{gprime_eqn}.
  The straight-line form implies that
  $S(\delta)\sim\e^{\alpha(\delta-\delta_c)^{-\beta}}$.  A least-squares
  fit (solid line) gives $\beta=0.499\pm0.001$, and we conjecture that the exact
  result is $\beta=\frac12$.}
\label{lnregress}
\end{figure}

To investigate the order of the transition in our model, we numerically
integrated Eq.~\eref{gprime_eqn} near $\delta=\frac18$.  The log-log plot
in Fig.~\ref{lnregress}a suggests that the size of the giant component
approaches zero faster than a power law as $\delta$ approaches $\delta_c$.
In Fig.~\ref{lnregress}b, we take an additional logarithm and plot
$\log(-\log(S))$ against $\log(\delta-\delta_c)$.  The resulting points now
appear to fall on a straight line as we approach the transition, indicating
that the size of the giant component is well approximated by a function of
the form

\begin{equation}
S(\delta) \sim \e^{\alpha(\delta-\delta_c)^{-\beta}}
               \qquad\mbox{as $\delta\to\delta_c$},
\label{Scrit}
\end{equation}
where the straight line in the figure implies that the leading constant is
unity.  The form of Eq.~\eref{Scrit} suggests that the phase transition is
in fact of infinite order, since all derivatives vanish at~$\delta_c$.  If
true, this would be an interesting result.  Most known phase transitions of
infinite order are of the Kosterlitz--Thouless
type~\cite{bere70,kost73,itoi99}, i.e.,~they arise in models that can be
mapped to a two-dimensional Coulomb gas, such as the two-dimensional
classical XY model or the nonlinear $\sigma$-model.  Because there is no
obvious mapping from our system to a two-dimensional one, it seems likely
that the transition here is produced by another mechanism.

A least-squares fit to the data in Fig.~\ref{lnregress} gives
$\alpha=-1.132\pm0.008$ and $\beta=0.499\pm0.001$.  We conjecture that in
fact $\beta$ is exactly equal to $\frac12$, and hence that the
appropriate asymptotic form for $S$ is
$S(\delta)\sim\e^{\alpha/\sqrt{\delta-\delta_c}}$.

\section{Degree correlations}
\label{corr}
The results of the previous sections indicate that the behavior of grown
random graphs is distinctly different from that of static random graphs.
Why should this be?  What is it about a grown graph that makes the giant
component form at a lower density of edges than in the corresponding
static graph?  The crucial difference seems to be that in the grown
graph some vertices are older than others, having been added to the
graph earlier, whereas in the static graph, all vertices are
added at the same time.  The older vertices in the grown graph have a
higher probability of being connected to one another, since they
co-existed earlier, and hence had the opportunity of developing
connections at a time when the graph was relatively small.  Thus the
graph has a ``core'' of old vertices in which there is a higher than
average density of edges.  Because of this core, the giant component
forms more readily than in a graph whose edges are uniformly
distributed.  On the other hand, as $\delta$ increases, the size of
the giant component in the growing graph increases more slowly than in
the static graph, since low-degree vertices remain connected only to
one another, rather than joining the giant component.

To demonstrate the effect of the differing ages of vertices, we now examine
correlations between the degrees of connected vertices in the growing
graph.  Since older vertices tend also to be vertices of higher degree, we
can test our hypothesis about the existence of a core by determining
whether vertices of high degree tend to be connected to one another more
often than one would expect in a static random graph.

We define $E_{kl}(t)$ to be the number of edges with a vertex of degree $k$
at one end and a vertex of degree $l$ at the other, at time~$t$.  This is
the discrete-time version of a quantity recently introduced by
Krapivsky\etal~\cite{krap00b} in the study of preferential attachment
models.  There are three possible processes that increase the value of
$E_{kl}$ as our network grows: (1)~a vertex of degree $k-1$, already
connected to a vertex of degree~$l$, is chosen for attachment to third
vertex of any degree; (2)~the same process with $k$ and $l$ reversed;
(3)~two vertices with degrees $k-1$ and $l-1$ vertices are chosen for
connection to one another.  Similarly there are two possible processes
that decrease $E_{kl}$: (1)~a vertex of degree~$k$ that is attached to a
vertex of degree~$l$ gains an additional edge; (2)~the same process with
$k$ and $l$ reversed.  As in the derivation of the component size
distribution, we are interested in the behavior of the graph only in the
large system size limit, and thus we can safely neglect higher-order
processes such as an edge connecting two previously connected vertices.

Given the processes described above, the difference equations governing
the time evolution of $E_{kl}$ are:
\begin{eqnarray}
\label{Edge_eqn}
E_{kl}(t+1) &=& E_{kl}(t) \nonumber\\
            & & + 2\delta\left(p_{k-1}\frac{E_{k-1,l}(t)}{d_{k-1}(t)} +
                p_{l-1}\frac{E_{k,l-1}(t)}{d_{l-1}(t)}\right)\nonumber\\
            & & + 2\delta p_{k-1}p_{l-1}\nonumber\\
            & & - 2\delta\left(p_{k}\frac{E_{kl}(t)}{d_{k}(t)}
                + p_{l}\frac{E_{kl}(t)}{d_{l}(t)}\right)
\end{eqnarray}
where the second and third lines correspond to the three processes above by
which $E_{kl}$ is increased, and the fourth to the processes by which it is
decreased.  As before, $p_k$ is the probability that a randomly chosen
vertex has degree $k$, and $d_k$ is the expected number of degree-$k$
vertices.

Note that $E_{kl}(t)$ satisfies $\sum_{kl} E_{kl}(t) = 2\delta t$,
which suggests that the large-$t$ solution will have the form $E_{kl}(t) =
2\delta t e_{k,l}$, where $e_{kl}$ is asymptotically independent of time.
Making this substitution and solving for $e_{kl}$ yields
\begin{equation}
\label{edge_eqn}
e_{kl} = \frac{2\delta}{1+4\delta}(e_{k-1,l}+e_{k,l-1})
         + \frac{p_{k-1}p_{l-1}}{1+4\delta}.
\label{ekl}
\end{equation}

To quantify the tendency for edges to connect vertices of like degree, we
compute the degree correlation coefficient:
\begin{equation}
\rho = \frac{c}{\sigma^2}.
\label{defsrho}
\end{equation}
where
\begin{equation}
\sigma^2 = {\sum_k (k-\mu)^2 k p_k \over \sum_l l p_l}
\label{defss1}
\end{equation}
is the variance of the distribution of vertex degree at either end of a
randomly chosen edge, and
\begin{equation}
c = \sum_{kl} (k-\mu)(l-\mu) e_{kl}
\label{defss2}
\end{equation}
is the covariance between vertex degrees at both ends.
In these expressions
\begin{equation}
\mu = {\sum_k k^2 p_k\over\sum_k k p_k}
\label{defsmu}
\end{equation}
is the average degree of a vertex at the end of a randomly chosen edge.

\begin{figure}
\begin{center}
\psfig{figure=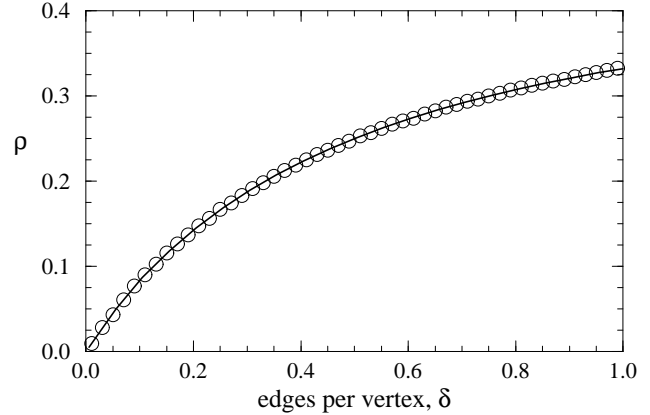,width=3.3in}
\end{center}
\caption{The correlation coefficient for the degrees of connected
  vertices in a randomly grown graph.  The solid line is the analytic
  result, Eq.~\eref{rho_eqn}, and the open circles are numerical results
  from simulations of the growth model for 10$^6$ time steps, averaged over
  25 realizations for each value of~$\delta$. }
\label{rho_fig}
\end{figure}

Substituting Eqs.~\eref{p_k_soln}, \eref{ekl}
and~(\ref{defss1}--\ref{defsmu}) into Eq.~\eref{defsrho}, we find
\begin{equation}
\rho = \frac{\sum_{kl} [k-(1+4\delta)] [l-(1+4\delta)] e_{kl}}
       {4\delta(1+2\delta)}.
\label{rho_eqn}
\end{equation}
In Fig~\ref{rho_fig} (solid line) we show this value for $\rho$ as a
function of $\delta$, where the values of the quantities $e_{kl}$ are
derived from numerical iteration of Eq.~\eref{ekl}.  On the same figure we
show results for the same quantity from direct simulations of the growth
model (circles), and the two calculations are clearly in good agreement.

The analogous correlation coefficient in the static graph is identically
zero---the two ends of any given edge are placed independently of one
another.  So the positive value of $\rho$ in the grown graph indicates
that high-degree vertices attach to other high-degree vertices more often
than they would in a static random graph with the same degree
distribution, and suggests that our supposition about the observed
differences between grown and static graphs was in fact correct.  

\section{Conclusions}
\label{concs}
We have introduced and analyzed a model of a growing network.  The model is
simple enough that many of its properties are exactly solvable, yet it
shows a number of non-trivial behaviors.  The model demonstrates that even
in the absence of preferential attachment, the fact that a network is
grown, rather than created as a complete entity, leaves an easily
identifiable signature in the network topology.

The size of the giant component in a graph has been likened to the strongly
connected component of the World Wide Web (another growing
network)~\cite{brod00,albe00b,cohe00,call00,cohe01}.  In this context it is
interesting to note that it takes only half as many edges to produce a
giant component in the grown graph than in the corresponding static one.
Put another way, the giant component in the grown graph is more robust to
random edge deletion; twice as many edges would have to be removed from it
to destroy its giant component.  It is possible that a similar process
helps large growing graphs like the Web achieve and maintain a high level
of overall connectivity even for low edge densities.

We have also shown that there is a positive correlation between the degrees
of connected vertices in our model, or equivalently that highly connected
vertices are preferentially connected to one another.  Similar correlations
have been observed previously in preferential attachment
models~\cite{krap00b}.  However, our results can be interpreted in a new
way---in our case the degree correlation appears to force the early
emergence of the giant component, and thus alters the component size
distribution in a graph that is otherwise random.

A number of interesting questions remain to be answered about the model
described here.  In particular, although we have an exact differential
equation for the generating function of component sizes, we have no exact
solution for this equation, and hence no closed form result for the
distribution of component sizes.  We also have at present only numerical
evidence that the phase transition is of infinite order.  Further work on
these and other questions would help to shed light on the unusual behavior
of this model.

This work was supported in part by the National Science Foundation, the
Department of Defense, and the Electric Power Research Institute.

\end{document}